\documentstyle[12pt,aps,prd,epsf,amsfonts]{revtex}

\def\rpm{R_p \hspace{-0.8em}/\;\:}
\def\rp{$R_p \hspace{-1em}/\ \  $}
\begin{document}


\title{Neutrino Oscillations and 
R-parity Violating Supersymmetry
\\[2ex]}

\author{
O. Haug\footnote{oliver.haug@uni-tuebingen.de}$^{1}$,
J. D. Vergados$^{1,2}$,
Amand Faessler$^{1}$
and S. Kovalenko$^{1,3}$\\[4ex]}

\address{
$^1$Institut f\"ur Theoretische Physik, Universit\"at T\"ubingen,
Auf der Morgenstelle 14, \\
D-72076 T\"ubingen, Germany\\
$^2$Theoretical Physics Division, Ioannina University, Ioannina, Greece\\
$^3$Joint Institute for Nuclear Research, Dubna, Russia}

\maketitle
\begin{center}
{\bf Abstract}
\end{center}
\begin{abstract}
Using the neutrino oscillations and neutrinoless
double beta decay experimental data we reconstructed an upper
limit for the three generation neutrino mass matrix.
We compared this matrix with the predictions
of the minimal supersymmetric(SUSY) model with R-parity violation(\rp)
and extracted stringent limits on trilinear \rp coupling constants
$\lambda_{i33}, \lambda'_{i33}$.
Introducing an additional $U(1)_X$ flavor symmetry
which had been successful in explaining
the mass hierarchy of quarks and charged leptons we were able
to relate various \rp parameters. In this model we found
a unique scenario for the neutrino masses and the \rp couplings
compatible with the neutrino oscillation data. Then we derived
predictions for certain experimentally interesting observables.
\end{abstract}
\pacs{12.60.Jv, 11.30.Er, 11.30.Fs,  23.40.Bw}
\newpage
\section{Introduction}
\label{intro}
It is a common belief that the existence of neutrino oscillations point to 
physics beyond the standard model $(SM)$.
Recent Super-Kamiokande results strongly
support the existence of neutrino oscillations \cite{fukuda:98}
by the observation of the zenith-angle
dependence of the high energy atmospheric $\nu_{\mu}$ events.
Other hints for this phenomenon come
from the solar neutrinos \cite{homestake:98,sage:98} and
the accelerator LSND \cite{louis:98}-\cite{lsnd:98}
neutrino oscillation experiment.

The neutrino data were extensively used for testing
various models of physics beyond the standard model \cite{Valle:99D}.
Recently there was a growing interest in the description
of neutrino properties in supersymmetric models with
R-parity violation (R-parity violating Minimal Supersymmetric Standard Model 
\rp MSSM). It was realized a few years ago
that the \rp MSSM framework is rather adequate for this purposes.
A non-trivial Majorana neutrino mass matrix is a generic
feature of the \rp MSSM as a consequence of the lepton violating
\rp couplings \cite{nu-LH,bgnn96}.

In the present paper we are studying the impact of
neutrino oscillation data on the three family neutrino
mass matrix and on the flavor structure of the \rp MSSM couplings.

There exists a controversy, whether
a three neutrino family scenario is able to accommodate
all these data or an additional fourth light sterile neutrino
must be included in the theory \cite{gnpv98}.
Recently it was argued that three neutrinos are enough for
a reasonable description \cite{scheck:98}-\cite{ohlsson:99} of all
the above cited neutrino oscillation data.
A especially good fit to the data was obtained
by taking out the LSND points from the analysis. This is motivated
by the opinion that the LSND result needs an independent confirmation.

Here we accept the three family
neutrino scenario. In section \ref{phen} we start
with consideration of the constraints imposed on
this scenario \cite{scheck:98}-\cite{ohlsson:99}
by the neutrino oscillation data and show that in this framework one needs
additional information on the overall neutrino mass scale in order to
determine the neutrino mass matrix. Towards this end we use the neutrinoless
double beta decay ($0\nu\beta\beta$) experimental constraints on the average neutrino
mass $\langle m_{\nu}\rangle$ and reconstruct the three family neutrino mass
matrix with the maximal entries allowed by these data.

In section \ref{model} we consider
the three family neutrino mass matrix in
the \rp MSSM \cite{nilles:84}. We allow a most general case of explicit
$R$-parity violation in the superpotential and the soft
SUSY breaking sector \cite{Rp-rev,barbier:98} taking into account
both the trilinear and the bilinear \rp terms.
In this model the neutrinos acquire masses
at the electroweak scale via tree level neutrino-neutralino
mixing as well as via 1-loop corrections \cite{nu-LH,bgnn96}.
We compare the total 1-loop three neutrino family mass matrix of
the \rp MSSM with the maximal mass matrix derived in section
\ref{phen}
and extract stringent constraints on the trilinear \rp couplings
$\lambda_{i33},\lambda'_{i33}$. Similar constraints on the bilinear \rp
parameters and products of certain trilinear \rp couplings
from the neutrino oscillation data were previously derived in ref.
\cite{bednyakov:98}-\cite{chung:98}.

It is known that the predictive power of the  \rp MSSM is quite weak
due to the presence of many free parameters. In section \ref{hierarchy}
we consider a model based on the presently popular idea of the horizontal
$U(1)_X$ flavor symmetry. Being imposed on the \rp MSSM this symmetry
relates many parameters and allows one to very successfully describe
the quark and the charged lepton masses and mixing angles
\cite{ibanez:94,lola:98}. The \rp couplings are also subject to $U(1)_X$
symmetry relations. These relations restrict the \rp MSSM neutrino
mass matrix so that the overall neutrino mass scale becomes fixed
only by the neutrino oscillation data. We find a unique solution
for the neutrino masses and the trilinear \rp couplings for every 
oscillation analysis.
On this basis we predict the average neutrino mass $\langle m_{\nu}\rangle$
and shortly discuss prospects for the future $0\nu\beta\beta$-decay experiments.

\section{Neutrino oscillations and neutrino mass matrix.
Phenomenological treatment}
\label{phen}
Neutrino oscillations can occur if neutrinos have a non vanishing rest mass
and
their weak eigenstates $|\nu^o_{\alpha} \rangle $, $\alpha = e,\mu,\tau$,
do not coincide with the mass eigenstates $|\nu_{i} \rangle$, $i=1,2,3$.
The unitary mixing matrix $U$ that relates the weak and mass eigenstates can
be parameterized in the three family scenario by the three angles
$\theta_{12},\theta_{13},\theta_{23}$.
Assuming that $CP$-violation is negligible one gets
\begin{eqnarray}
U & = &
\left(
\begin{array}{ccc}
c_{12} c_{13} & s_{12} c_{13} & s_{13} \\
-s_{12} c_{23}-c_{12}s_{23}s_{13} & c_{12}c_{23}-s_{12}s_{23}s_{13}&
s_{23}c_{13} \\
s_{12}s_{23}-c_{12}c_{23}s_{13} & - c_{12}s_{23}-s_{12}c_{23}s_{13}&
c_{23}c_{13}
\end{array}
\right),
\end{eqnarray}
where $s_{ij}$ and $c_{ij}$ stand for $\sin\theta_{ij}$ and
$\cos\theta_{ij}$ respectively.
Then a neutrino produced in the weak eigenstate
$|\nu^o_{\alpha}\rangle$ changes its flavor content when
propagating in space. The probability of finding a neutrino
produced in the flavor state $\alpha$ at a given distance $L$
from the production point with the energy $E$ in the flavor state
$\beta$ is given by
\begin{eqnarray}
P(\alpha \rightarrow \beta)
& = &
\delta_{\alpha , \beta} - 4 \sum_{i < j = 1}^{3}
U_{\alpha i} U_{\beta i} U_{\alpha j} U_{\beta j}
\sin^2
\left[
\frac{\Delta m^2_{ij} L}{4 E}
\right].
\label{prob}
\end{eqnarray}
Here $\Delta m^2_{ij} \equiv \left| m_i^2 - m_j^2 \right| $
is the difference of the squared masses of the neutrino mass
eigenstates $i$ and $j$. The phenomenological neutrino mass
matrix ${\cal{M}}^{ph}$ in the flavor space is connected to
the physical neutrino masses $m_i$ by the mixing matrix $U$
as follows
\begin{eqnarray}
\label{diag}
  {\cal{M}}^{ph} & = & U\cdot diag ( m_1,m_2,m_3 )\cdot U^T.
\end{eqnarray}
Using eq. (\ref{prob}) one can extract
from the oscillation experiments the mixing angles
$\theta_{ij}$ and the squared mass differences $\Delta m^2_{ij}$.
This information is not sufficient for the restoration of the neutrino
mass matrix ${\cal{M}}^{ph}$.
The overall mass scale as well as the $CP$ eigenvalues
$\zeta^{(i)}_{CP}$ of the neutrino mass eigenstates
$(+1$ or -1 in our model$)$
remain undetermined. To fix this ambiguities one needs additional
experimental information other than the neutrino oscillation data.
This information exists in a form of upper limits on the neutrino
masses or their combinations. Experiments, measuring the neutrino masses
directly, offer at present too weak limits, leaving the neutrino mass matrix
very uncertain.
Much better result can be achieved using the neutrinoless
double beta decay ($0\nu\beta\beta$) constraints on the generation
average Majorana electron neutrino mass
\begin{eqnarray}
  \langle m_\nu\rangle & = & \sum_{i} m_{i}\zeta^{(i)}_{CP}
  \left(U_{ei}\right)^2.
\end{eqnarray}
From the currently best experimental limit on
$0\nu\beta\beta$-decay half-life of $^{76}$Ge
\cite{hdmo97}
\mbox{$
T_{1/2}^{{0\nu\beta\beta}}(0^+ \rightarrow 0^+)
\hskip2mm \geq \hskip2mm
1.1 \times 10^{25}~\mbox{years} \ \ (90\% \ \mbox{C.L.})
$}
one obtains $\langle m_\nu\rangle < 0.62$ eV
\cite{faessler:98,simkovic:99}.


Now with this additional input limit
we can find the maximal allowed values
for the matrix elements $m^{max}_{ij}$ of
the neutrino mass matrix. In our numerical analysis we are
searching for these maximal values over the whole allowed mass
parameter space. In doing this we take care of all the possible
$CP$-phases of the neutrino mass eigenstates.
The resulting absolute values of the matrix elements of this
"maximal" neutrino mass matrix are
\begin{eqnarray}
\label{max-numat}
|m^{max}| & = &
 \left(
 \begin{array}{ccc}
  .60 & .97 & .85\\
  .97 & .76 & .80\\
  .85 & .80 & 1.17
 \end{array}
 \right) \mbox{eV}.
\end{eqnarray}
Here we used the results of
the phenomenological analysis of
the neutrino oscillation data (including the LSND data) made in refs.
\cite{scheck:98}-\cite{ohlsson:99}.
In eq. (\ref{max-numat}) the worst case of the weakest bounds is given.
This "maximal" neutrino mass matrix can be used to
test various theoretical approaches and allows one to extract limits on
certain fundamental parameters. Below we are studying in
this respect the \rp MSSM and find new limits on the \rp parameters.
\section{Neutrino masses in $R_p \hspace{-1.0em}/\ \  $ MSSM}
\label{model}
The MSSM is the minimal supersymmetric extension of the SM.
%
Assuming that R-parity, defined as $R_P=(-1)^{3B+L+2S}$
($B$, $L$ and $S$ are the baryon, lepton numbers and the spin),
is conserved one ends up with the superpotential
\begin{eqnarray}
  \label{Rp-sup}
         W_{R_p} = \lambda^E_{ij} H_1 L_i E^c_j +
                   \lambda^D_{ij} H_1 Q_i D^c_j +
                   \lambda^U_{ij} H_2 Q_i U^c_j + \mu H_1 H_2.
\end{eqnarray}
Here $L$,  $Q$  stand for lepton and quark
doublet left-handed superfields while $ E^c, \  U^c,\   D^c$
for lepton and  up, down quark singlet  superfields;
$H_1$ and $H_2$ are the Higgs doublet superfields
with a weak hypercharge $Y=-1, \ +1$, respectively.

In the MSSM with conserved R-parity neutrinos remain massless
particles as in the SM. This follows from the fact that in this framework
there is no room for gauge invariant neutrino mass terms of either
Dirac or Majorana type.

Since R-parity conservation has no robust
theoretical motivation one may accept an extended framework of
the MSSM with R-parity non-conservation (\rp MSSM).
In this case the superpotential
$W$ acquires additional \rp terms \cite{nilles:84}
\begin{eqnarray}
  \label{rpsuperpotential}
  W_{\rpm} & = & \lambda_{ijk}L_{i}L_{j}E^c_{k}
  + \lambda^{\prime}_{ijk}L_{i}Q_{j}D^c_{k}
  +\mu_j L_{j}H_2
  + \lambda^{\prime \prime}_{ijk} U^c_{i}D^c_{j}D^c_{k},
\end{eqnarray}
so that the \rp MSSM superpotential is $W = W_{R_p} + W_{\rpm}$.

In general, R-parity is also broken in the "soft" SUSY breaking sector by
the scalar potential terms
\begin{eqnarray}
  \label{vsoft}
  V_{\rpm}^{soft} & = & \Lambda_{ijk}\tilde{L}_i\tilde{L}_j\tilde{E}^c_k
  +\Lambda^{\prime}_{ilk}\tilde{L}_{i}\tilde{Q}_{j}\tilde{D}^c_{k}
  +\Lambda^{\prime \prime}_{ijk}\tilde{U}^c_{i}\tilde{D}^c_{j}\tilde{D}^c_{k}
  +\tilde{\mu}^2_{2j}\tilde{L}_j H_2
  +\tilde{\mu}^2_{1j}\tilde{L}_j H^{\dagger}_1
  +H.c.   
\end{eqnarray}
The terms in (\ref{rpsuperpotential}) and
(\ref{vsoft})  break lepton and baryon number conservation.
The tilde indicates that one includes only supersymmetric fields. 
To prevent fast proton decay one may assume
$\lambda^{\prime \prime} = \Lambda^{\prime \prime} = 0$ that
is commonly expected as a consequence of certain
symmetry like baryon parity \cite{B-parity}.

The R-parity conserving part of the soft SUSY breaking
sector includes the scalar field interactions
\begin{eqnarray}\label{V_Soft}
V_{R_p}^{soft}= \sum_{i=scalars}^{}  m_{i}^{2} |\phi_i|^2 +
\lambda^E A^E L H_1 \tilde L \tilde{E}^c + \lambda^D A^D H_1 \tilde Q \tilde{D}^c +&&
\\  \nonumber
+ \lambda^U A^U H_2 \tilde Q \tilde{U}^c  +
\mu B H_1 H_2 + \mbox{ H.c.}&&
\end{eqnarray}
and the "soft"  gaugino mass terms
\begin{eqnarray}\label{M_soft}
{\cal L}_{GM}\  = \ - \frac{1}{2}\left[M_{1}^{} \tilde B \tilde B +
 M_{2}^{} \tilde W^k \tilde W^k  + M_{3}^{} \tilde g^a \tilde g^a\right]
 -   \mbox{ H.c.}
\end{eqnarray}
Here $M_{3,2,1}$ stand for the "soft" masses of the $SU(3)\times
SU(2)\times U(1)$ gauginos $\tilde g, \tilde W, \tilde B$ while $m_i$
denote the masses of the scalar fields.

In the above sketched framework of the \rp MSSM
neutrinos are, in general, massive. Generically, one can
distinguish the following three contributions to the neutrino masses:
\begin{itemize}
\item Tree-level contribution:\\
The bilinear terms in eqs. (\ref{rpsuperpotential}),
(\ref{vsoft}) lead to terms in the scalar
potential linear in the sneutrino fields and thereby a non-vanishing
vacuum expectation values(VEVs) of these fields
$\langle\tilde\nu_i\rangle\neq 0$. This leads to a mass
term for the neutrinos by mixing with
the gaugino fields $\tilde{B}^{0}$ and $\tilde{W}^{3}$.
The term
$\mu_{i} L_{i} H_{2}$ in equation (\ref{rpsuperpotential})
gives an additional mass term from the mixing of neutrinos with the neutral
Higgsino fields $\tilde{H}^{0}_1,\tilde{H}^{0}_2$.
The so generated non-trivial $7 \times 7$ mass matrix in the
basis
$(\nu_e,\nu_{\mu},\nu_{\tau},\tilde{B}^{0}, \tilde{W}^{3},\tilde{H}^0_1,
\tilde{H}^{0}_{2})$ can be brought into a block diagonal form
\cite{nowakowski:96}
and an effective neutrino mass matrix at tree level
$M^{tree}$ can be extracted. In leading order in implicitly small
expansion parameters $\langle\tilde\nu_i\rangle/M_{SUSY}, \mu_i/M_{SUSY}$
one gets the expression
\cite{nowakowski:96}:
\begin{eqnarray}
  \label{mtree}
  {M}^{tree}_{\alpha \beta} & = & {\cal Z}_1
  \Lambda_{\alpha} \Lambda_{\beta}, \ \ \
  \Lambda_{\alpha} = \mu \langle \tilde{\nu}_{\alpha}\rangle
  -\langle H_1 \rangle \mu_{\alpha}, \\
  {\cal Z}_1 & = & g^2_2
  \left|
    \frac{M_1+\tan^2 \theta_W M_2}
    {4 ( \sin 2\beta\  M^2_W\  \mu\ (M_1+\tan^2\theta_W M_2) -
M_1\  M_2\ \mu^2)}
  \right| \nonumber
\end{eqnarray}

Here $g_2$ is the $SU(2)$ gauge coupling and
$\tan\beta = \langle H_2^0\rangle/\langle H_1^0\rangle$.
The "soft" SUSY breaking gaugino masses $M_{2,1}$ and the superpotential
parameter $\mu$ are usually assumed to be not too far from the characteristic
SUSY breaking scale $M_{SUSY}\sim 100$GeV.

\item $q\tilde{q}$-loop contribution:\\
Another contribution to the neutrino masses arises due to
the quark-squark self-energy loops, coming from the term
$\lambda^{\prime}_{ijk}L_{i}Q_{j}D^c_{k}$ in the R-parity
violating superpotential in eq. (\ref{rpsuperpotential}).
The corresponding diagram is shown in fig. 1(a) and its
contribution to the neutrino mass matrix is given by
\begin{eqnarray}
  M^{q\tilde{q}}_{ij} & \simeq &
  \sum_{k,l,m}
  \frac{3{\lambda}^{\prime}_{ikl} \lambda^{\prime}_{jmn}}
  {8 \pi^2}
  \frac{M^d_{kn} M^d_{ml}(A^D_{ml}+\mu \tan \beta)}{\widetilde{m}^2_{d_l}}
  \nonumber
  \\
  \label{mqsq}
  & \sim &
  \frac{3{\lambda}^{\prime}_{i33} \lambda^{\prime}_{j33}}{8 \pi^2}
  \frac{m^2_{b}(A^D_{b}+\mu \tan \beta)}{\widetilde{m}^2_{b}}
  \equiv {\cal Z}_2\lambda^{\prime}_{i33} \lambda^{\prime}_{j33}.
\end{eqnarray}
In the numerical analysis
we assumed the down quark mass matrix $M^d$ to be diagonal and
keep only the dominant contribution of the $b\tilde b$-loop which is 
proportional
to $m_b^2$.
The dominance of the heaviest internal fermion line holds for
non-hierarchical $\lambda'_{ijk}$ and $A^D_{ij}$ in a sense
that they do not strongly grow with increasing generation
indices. We also assume these quantities to be real.
The factor 3 appears from summation over the (s)quark colors in the loop.

\item $l\tilde{l}$-loop contribution:\\
The last contribution to the neutrino mass matrix at
the 1-loop level is given by the diagram in fig. 1(b).
It is induced by the $L_i L_j E^c_k$ term in eq.  $(\ref{rpsuperpotential})$.
It has the same structure as the above discussed $q\tilde q$-loop.
The $l\tilde{l}$-loop contribution to the neutrino mass matrix reads
\begin{eqnarray}
  M^{l\tilde{l}}_{ij} & \simeq &
  \sum_{k,l,m}
  \frac{{\lambda}_{ikl} \lambda_{jmn}}
  {8 \pi^2}
  \frac{M^e_{kn} M^e_{ml}(A^E_{ml}+\mu \tan \beta)}{\widetilde{m}^2_{e_l}}
  \nonumber
  \\
  \label{mlsl}
  & \sim &
  \frac{{\lambda}_{i33} \lambda_{j33}}{8 \pi^2}
  \frac{m^2_{\tau}(A^E_{\tau}+\mu \tan \beta)}{\widetilde{m}^2_{\tau}}
  \equiv {\cal Z}_3 \lambda_{i33} \lambda_{j33}.
\end{eqnarray}
Here,
as in the case of the $q\tilde{q}$-loop, we assumed absence of
the generation index hierarchy in $\lambda_{ijk}$ and $A^E_{ij}$ and
the charged lepton mass matrix $M^e$ to be diagonal. For
the numerical analysis we kept only
the dominant $\tau\tilde\tau$-loop contribution.

The quantities $\widetilde{m}^2_d$ and $\widetilde{m}^2_e$
in eqs. (\ref{mqsq}), (\ref{mlsl}) denote
the left-right averaged square of squark and slepton masses respectively.
\end{itemize}

Summarizing, we write down the 1-loop level neutrino mass matrix
${\cal M}^{\nu}$ in the \rp MSSM as
\begin{eqnarray}
  \label{MT}
{\cal M}^{\nu} & = & M^{tree} + M^{q\tilde{q}} + M^{l\tilde{l}}.
\end{eqnarray}

From eqs. (\ref{mtree})-(\ref{mlsl}) one sees that this matrix is of the from
\begin{eqnarray}
  \label{MTT}
M_{ij}^{\nu}=\sum_{k=1}^3 a^k_i a^k_j
,
\end{eqnarray}
with the three terms in the sum corresponding to
the three terms in eq. (\ref{MT}) built of the three
different 3-dimensional vectors $\vec{a}^k$. A matrix
with such a  structure has no eigenvector $\vec x^{(0)}$
with eigenvalue zero and therefore all the three neutrinos
have non-zero masses. This follows from
the fact that the zero-eigenmass condition $M^{\nu}\cdot \vec x^{(0)} = 0$
requires the vector $\vec x^{(0)}$ to be simultaneously
orthogonal to the three different vectors $\vec a^k$ ($k=1,2,3$)
which is impossible in the 3-dimensional space. The same arguments show
that neglecting one or two of the terms in
eqs. (\ref{MT}), (\ref{MTT}) results in one or two zero-mass neutrino
states respectively. Thus in order to keep all the neutrinos massive we
retain all three terms in eq. (\ref{MT}).

In section \ref{phen} we extracted
the limits on the matrix elements of
the neutrino mass matrix in the three family scenario.
Now we can translate these limits to limits on
the trilinear couplings of the \rp MSSM. Towards this end
it is enough to analyze only the diagonal matrix elements.
Note that in case when all the three normalization
factors ${\cal Z}_{1,2,3}$ in eqs. (\ref{mtree})-(\ref{mlsl})
have the same sign no compensations occur between different
terms contributing to these matrix elements since they would be a sum
of positive(negative) terms. This would allow one to get
an upper bound not only for the sum but also for each term of the sum 
separately. 
As follows from the Renormalization Group Equation
analysis \cite{castano:94} this condition is satisfied
for a quite wide region of the MSSM parameter space but not
everywhere. In our order of magnitude
analysis it is enough to assume that there are no
large compensations between the different terms.
For the MSSM parameters we take
$A \simeq \mu \simeq m_{\tilde{b}} \simeq m_{\tilde{\tau}} \simeq M_{SUSY}$
and $\tan \beta = 1$.
The characteristic SUSY breaking mass scale $M_{SUSY}$ is usually
assumed to vary in the interval 100 GeV $\leq $ $M_{SUSY}$ $\le $ 1 TeV
motivated by non-observation of the superparticles and by the "naturalness"
arguments.
With this choice of parameters
we obtained the upper bounds for the trilinear \rp couplings shown in table
\ref{table1}. There we also display the existing bounds for this
parameters \cite{rakshit:98}.
One sees from the table \ref{table1} that our limits are more
stringent than the previously known ones.

\section{\rp MSSM with family dependent $U(1)_X$ symmetry}
%
%
\label{hierarchy}
Now suppose that the \rp MSSM Lagrangian is invariant under
the family dependent $U(1)_X$ symmetry.
Presently this is a popular idea which allows one to predict
the hierarchical structure of the charged
fermion mass matrices and the fermion masses are in agreement with
low-energy phenomenology \cite{ibanez:94,lola:98}. In this
\rp MSSM$\times U(1)_X$-model the trilinear \rp couplings are forbidden
by the $U(1)_X$ symmetry and appear when it is spontaneously broken.
These couplings can be generated
by the effective operators
\begin{eqnarray}
  \label{eff-oper}
L_{i}L_{j}E^c_{k} \left(\frac{\theta}{M_X}\right)^{l_i + l_j + e_k},
\ \ \ \ \ \ \ \ \ \ \ \ \ \
L_{i}Q_{j}D^c_{k}\left(\frac{\theta}{M_X}\right)^{l_i + q_j + d_k},
\end{eqnarray}
existing in the $U(1)_X$ symmetric phase and originating from
the underlying theory at the large scale $M_X$. In eq. (\ref{eff-oper})
we denoted the $U(1)_X$ charges of the quark and lepton fields as
$l_i, q_j, d_k, e_k$.
The SM singlet field $\theta$ has the $U(1)_X$ charge -1 and, acquiring
the vacuum expectation value $\langle\theta\rangle \neq 0$, breaks
the $U(1)_X$ symmetry. As a result the effective operators
in eq. (\ref{eff-oper}) generate the following effective couplings
\cite{barbier:98,binetruy:98}
\begin{eqnarray}
  \label{tril1}
  \lambda_{ijk} & \sim & \epsilon^{\tilde{l}_i-\tilde{l}_0} \lambda^E_{jk}, \\
  \label{tril2}
  \lambda^{\prime}_{ijk} & \sim & \epsilon^{\tilde{l}_i-\tilde{l}_0 }
  \lambda^D_{jk}.
\end{eqnarray}
Here is $\epsilon = \langle\theta\rangle/M_0$
and $\tilde l_{i} = |l_{i} +  h_2|$ with $l_{i}, h_2$ (i=1,2,3)
being the $U(1)_X$ charges of the lepton and $H_2$ Higgs fields.
The parameter $\epsilon \approx 0.23$
and the relative $U(1)_X$ charges
\begin{eqnarray}
\label{charges}
|\tilde{l}_1 - \tilde{l}_3| = 4, \ |\tilde{l}_2 - \tilde{l}_3| = 1
\end{eqnarray}
were found in the analysis of the charged lepton and quark mass
matrices in refs. \cite{ibanez:94,lola:98}.
The remaining ambiguity in the flavor
independent quantity $\tilde{l}_0$ can be removed by taking ratios.

Note that the formulas in
eqs. (\ref{tril1}),(\ref{tril2})
are given in the field basis where the VEVs of the sneutrino
fields are zero  $\langle\tilde\nu_{1,2,3}\rangle = 0$.  Translation to this
specific basis is achieved by
the unitary rotation in the field subspace $L_{\alpha} = (L_i, H_1)$
\cite{barbier:98,binetruy:98}.

The trilinear lepton \rp couplings in
eq. $(\ref{tril1})$ must be antisymmetric
under interchange of the first two indices $i$ and $j$. 
This leads to the expression
\begin{eqnarray}
  \lambda_{ijk} & \sim & \frac{1}{2}
  \left(
  \epsilon^{\tilde{l}_i-\tilde{l}_0} \lambda^E_{jk}-
  \epsilon^{\tilde{l}_j-\tilde{l}_0} \lambda^E_{ik}
  \right).
\end{eqnarray}
The ratio of the $\lambda_{ijk}$ and $\lambda^{\prime}_{ijk}$ is then given
as
\begin{eqnarray}
  \frac{\lambda_{ijk}}{\lambda^{\prime}_{ijk}} & = &
  \frac{1}{2}
  \left(
  \frac{\lambda^E_{jk}}{\lambda^D_{jk}}-\epsilon^{\tilde{l}_i-\tilde{l}_j}
  \frac{\lambda^E_{ik}}{\lambda^D_{jk}}
  \right).
\end{eqnarray}
Approximating the ratio
${\lambda^{E}_{ik}}/{\lambda^D_{jk}}\sim -2$
(see ref \cite{binetruy:98}) we get for the $U(1)_X$ charges
given in eq. (\ref{charges}) the following \rp couplings
\begin{eqnarray}
  \vec{\lambda}^{\prime} & = & \lambda^{\prime}_{333}
  \left(
  \begin{array}{c}
  \epsilon^4 \\
  \epsilon \\
  1
  \end{array}
  \right),
\label{coupling1}
\\
  \vec{\lambda} & = & \lambda_0
  \left(
  \begin{array}{c}
  (\epsilon^4-1)\epsilon^4 \\
  (\epsilon-1)\epsilon \\
  0
  \end{array}
  \right),\lambda_0 \simeq \lambda^{\prime}_{333}.
\label{coupling2}
\end{eqnarray}
Here we denoted $\vec{\lambda}^{\prime}= \{\lambda^{\prime}_{i33}\},
\vec{\lambda}= \{\lambda_{i33}\}$.
In these eqs. remains only one free parameter $\lambda^{\prime}_{333}$.
Thus, the $U(1)_X$ symmetry allows us to dramatically reduce
the number of free parameters 
in the neutrino mass matrix given by eqs. (\ref{mtree})-(\ref{MT}).
Totally, in the \rp sector there
are only four free parameters:
the trilinear coupling $\lambda^{\prime}_{333}$ and
the three bilinear $\Lambda_{\alpha}$ parameters (see eq. (\ref{mtree})).
The latter three are also subject
to $U(1)_X$ constraints and under certain additional
assumption further reduction of free parameters is possible
\cite{barbier:98,binetruy:98}.
In our subsequent analysis we disregard these constraints
and keep the three $\Lambda_{1,2,3}$ quantities as free parameters,
avoiding additional assumptions.
We already pointed out, that from the neutrino oscillation
data alone one is able to fix the entries of the neutrino
mass matrix up to the overall mass scale and the sign ambiguities
which appear in solving the non-linear equations (\ref{prob}),(\ref{diag}).
As we have shown earlier in this section the \rp MSSM$\times U(1)_X$
neutrino mass matrix depends only on four effective parameters.
Now confronting the phenomenological
neutrino mass matrix ${\cal M}^{ph}$, derived from
the analysis of the neutrino data, with the \rp MSSM$\times U(1)_X$
mass matrix ${\cal M}^{\nu}$, given by eqs. (\ref{mtree})-(\ref{MT}),
we get a system of six linear independent equations
\begin{eqnarray}
\label{condformasses}
 {\cal M}^{ph} =:  {\cal M}^{\nu}
\end{eqnarray}
with five unknown quantities.
Solving this system of equations one can uniquely determine
the mass scale on the phenomenological side and in addition
the absolute values of the four theoretical parameters. The results for
the neutrino masses with the corresponding CP-phases $\zeta_{CP}$
and the family averaged Majorana neutrino mass
$\langle m_{\nu} \rangle  = \sum_{i}\zeta_i U^2_{ei} m_i$ are
given in table \ref{table2}.
The \rp parameters are shown in table \ref{table3}.
We made our numerical analysis under the assumptions discussed
at the end of section \ref{model}.
Our results are given for the four different sets of matrix elements
of the input matrix $ {\cal M}^{ph}$ found from the phenomenological analysis
of the neutrino data in refs. \cite{scheck:98}-\cite{ohlsson:99}.
Note, that for all the examined input sets
we found a hierarchical neutrino mass scenario.

The values of the bilinear parameters $\Lambda_{1,2,3}$
in table \ref{table3} are within the upper limits
found in refs. \cite{bednyakov:98,fks:98,fkkv:99}.
The same is true for the values of the trilinear \rp coupling
$\lambda^{\prime}_{333}$ in table \ref{table3} as well as for
all other couplings derived according to eqs. (\ref{coupling1}) and 
(\ref{coupling2}). They are not in conflict with
the limits we found in section \ref{model} and presented
in table \ref{table1}.

As seen from
table \ref{table2} our prediction for
the family averaged neutrino Majorana mass ranges in the interval
$|\langle m_{\nu} \rangle | \sim 0.01-0.05\mbox{eV}$.
This range is more
than one order of magnitude
below the existing limit of
$|\langle m_{\nu} \rangle | < 0.62$ eV\cite{faessler:98,simkovic:99}
extracted from the current $0\nu\beta\beta$-decay data \cite{hdmo97}.

\section{Summary}
We discussed the phenomenology of neutrino oscillations
and found upper limits on the entries of the three family
neutrino mass matrix. This we did by using the 
double beta decay constraints on the family average Majorana neutrino mass
$\langle m_{\nu}\rangle$.
The so derived "maximal" neutrino mass matrix
was used to test the flavor structure of the R-parity violating
sector of the \rp MSSM (R-parity violating Supersymmetric Standard Model).
Comparing
the theoretical 1-loop \rp MSSM neutrino mass matrix with
our phenomenological "maximal" matrix
we extracted new limits on the trilinear \rp coupling
constants $\lambda_{i33}$ and $\lambda^{\prime}_{i33}$.
These limits are more stringent than those existing in the literature.
For the $\lambda^{\prime}_{233}$,
$\lambda^{\prime}_{333}$ couplings are the new limits
an improvement of up to 3 orders in magnitude compared to the existing limits. 

As a next step we considered the \rp MSSM with
the family dependent horizontal $U(1)_X$ symmetry. The framework of
this \rp MSSM$\times U(1)_X$ model is rather restrictive and
allows one not only to set limits on the parameters from experimental
data but, what is more interesting, to derive predictions testable
in future experiments.
Accepting the $U(1)_X $ charge assignment previously obtained
in refs. \cite{ibanez:94} from the fit to the quark and charged lepton
masses and mixing angles we related various trilinear \rp coupling constants.
As a result only four parameters in the \rp MSSM$\times U(1)_X$ neutrino
mass matrix remained free. In this case we were able not only to determine
the upper limits but in addition the intervals of values for
these parameters from the existing neutrino oscillation without
any additional experimental information.
Moreover, we completely reconstructed in this framework the three
family neutrino mass matrix and give the predictions for the neutrino
masses as well as for the family average neutrino Majorana mass
$\langle m_{\nu} \rangle$.

Noteworthy, our prediction for the $\langle m_{\nu} \rangle$
lies in the interval $\sim 0.01-0.05 eV$. This is one order of magnitude
below the current experimental upper bound on this quantity.
However the next generation $0\nu \beta \beta$ experiments,
e.g. the GENIUS experiment \cite{GENIUS}, claim to be able to explore
this region of small values of the average Majorana neutrino masses.
Thus, we predict on the basis of the current neutrino oscillation data
a positive result of searching for $0\nu \beta \beta$-decay in the next
generation experiments with the sensitivity to $\langle m_{\nu} \rangle$
in the range of few tens milli-eV.
Since this prediction relies on
the \rp MSSM$\times U(1)_X$ observation of $0\nu \beta \beta$-decay
in this region would also indirectly support this model.

\section*{ACKNOWLEDGMENT}
OH wishes to thank Th. Gutsche for fruitful discussions.
JV is grateful for an award of the Humboldt Foundation.
OH was supported by the Graduiertenkolleg under contract DFG GRK 132/2.
\begin{appendix}
\end{appendix}




\newpage
\begin{figure}
\caption{The quark-squark (a)
and the lepton-slepton (b) 1-loop contribution
to the neutrino Majorana masses. The crosses on the lines denote
the left-right mixing.}
\label{fig1}
\end{figure}

\begin{table}
\caption{Upper limits for the trilinear R-parity violating couplings
derived from the neutrino oscillation and the neutrinoless double beta
decay data. Approximations are specified in \mbox{section \ref{model}.}}
\begin{tabular}{ccc}
        & new limit  & Existing bounds $($see $\cite{rakshit:98})$ \\ \hline
$\frac{\lambda_{133}}{M_{SUSY}/100 GeV}$&       1.7 $ \cdot 10^{-3}$  & 3 $\cdot 10^{-3}$\\
$\frac{\lambda_{233}}{M_{SUSY}/100 GeV}$&       1.9 $ \cdot 10^{-3}$  & 6 $\cdot 10^{-2}$\\
$\frac{\lambda^{\prime}_{133}}{M_{SUSY}/100 GeV}$&      3.8 $\cdot 10^{-4}$  &7 $\cdot 10^{-4}$\\
$\frac{\lambda^{\prime}_{233}}{M_{SUSY}/100 GeV}$& 4.3 $ \cdot 10^{-4}$ & .36 \\
$\frac{\lambda^{\prime}_{333}}{M_{SUSY}/100 GeV}$& 5.3 $ \cdot 10^{-4}$       & .48
\end{tabular}
\label{table1}
\end{table}

\begin{table}
\caption{The predictions of \rp MSSM with the $U(1)_X$ family symmetry
for the neutrino masses $m_i$ and the family average neutrino Majorana mass
$\langle m_{\nu} \rangle = \sum_{i}\zeta_{CP}^{(i)} U^2_{ei} m_i$.
Here $\zeta_{CP}^{(i)},i=1,2$ are the $CP$ phases
of the neutrino mass eigenstates. The assumptions are specified in
section \ref{hierarchy}. Different predictions correspond to
different input sets for the neutrino mixing angles $\theta_{ij}$
and $\Delta m^2_ij$ found from the neutrino oscillation data in
the papers cited in the last column.}
\begin{tabular}{ccccccc}
$m_1[eV]$ & $m_2[eV]$ & $m_3[eV]$ &$\zeta_{CP}^{(1)}$&$\zeta_{CP}^{(2)}$&
$\left| \langle m_{\nu} \rangle \right|$\\ \hline
.004    & .032  & .549  & +     & +     & .041	&\cite{scheck:98} \\
.018    & .036  & .549  & -     & +     & .045	&\cite{scheck:98} \\
.002    & .002  & .030  & +     & +     & .010	&\cite{fogli:99}  \\
.000    & .0224 &  .633 & -     & +     & .028	&\cite{thun:98}   \\
.019    & .026  & 1.054 & -     & +     & .009	&\cite{ohlsson:99}
\end{tabular}
\label{table2}
\end{table}

\begin{table}

\caption{The same as in table \ref{table2} but for the trilinear
\rp coupling $\lambda'_{333}$ and the bilinear \rp parameters $\Lambda_i$
defined in eq.  (\ref{mtree}).}
\begin{tabular}{ccccc}
$|\Lambda_1|[GeV^2]$&$|\Lambda_2|[GeV^2]$&$|\Lambda_3|[GeV^2]$
&$|\lambda^{\prime}_{333}/10^{-4}|$&
 \\ \hline
.008    & .012  & .019  & 2.1   &\cite{scheck:98} \\
.008    & .014  & .016  & 2.4   &\cite{scheck:98} \\
.004    & .004  & .002  & .7    &\cite{fogli:99}  \\
.006    & .013  & .021  & 2.1   &\cite{thun:98}   \\
.004    & .022  & .022  & 3.0   &\cite{ohlsson:99}
\end{tabular}
\label{table3}
\end{table}

\newpage

\centerline{Fig. 1}
\vskip 1cm
\centerline{\epsfbox{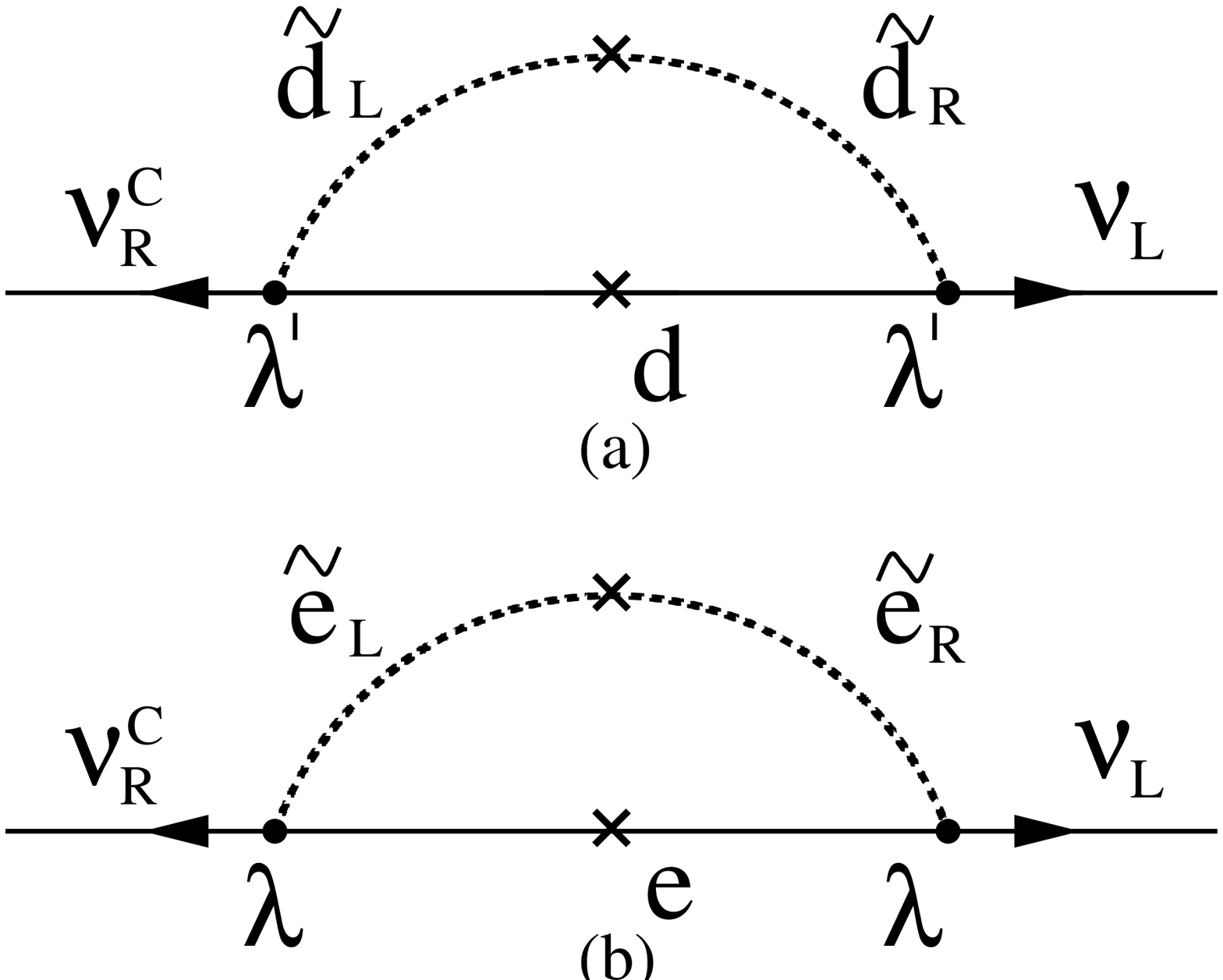}}
\end{document}